\documentclass[aps,prl,reprint,superscriptaddress]{revtex4-1}
\pdfoutput=1
\usepackage{easy-todo}
\usepackage{color}
\usepackage{graphicx}
\usepackage{amsmath}
\usepackage{amsfonts}
\usepackage{amssymb}
\usepackage{microtype}
\usepackage{verbatim}
\usepackage{bbold}
\usepackage{hyperref}
\usepackage{cleveref}
\allowdisplaybreaks
\usepackage{multirow}
\usepackage{bbm}
\usepackage{braket}
\hypersetup{
  colorlinks   = true, 
  urlcolor     = blue, 
  linkcolor    = blue, 
  citecolor   = blue 
}
\graphicspath{{./Figures/}}

\DeclareMathOperator{\tr}{tr}

\begin{document}

\author{Pieter W. Claeys}
\email{pc652@cam.ac.uk}
\affiliation{TCM Group, Cavendish Laboratory, University of Cambridge, Cambridge CB3 0HE, UK}

\author{Austen Lamacraft}
\affiliation{TCM Group, Cavendish Laboratory, University of Cambridge, Cambridge CB3 0HE, UK}

\title{Ergodic and non-ergodic dual-unitary quantum circuits \\ with arbitrary local Hilbert space dimension}

\begin{abstract}
Dual-unitary quantum circuits can be used to construct $1 +1$ dimensional lattice models for which dynamical correlations of local observables can be explicitly calculated. We show how to analytically construct classes of dual-unitary circuits with any desired level of (non-)ergodicity for any dimension of the local Hilbert space, and present analytical results for thermalization to an infinite-temperature Gibbs state (ergodic) and a generalized Gibbs ensemble (non-ergodic). It is shown how a tunable ergodicity-inducing perturbation can be added to a non-ergodic circuit without breaking dual-unitarity, leading to the appearance of prethermalization plateaux for local observables.
\end{abstract}

\maketitle
\emph{Introduction.} -- The dynamics of isolated systems under general unitary evolution remains one of the fundamental problems in many-body physics. 
Originating as a model for quantum computation \cite{Nielsen:2000aa}, unitary circuits can serve as a minimal model for the study of general unitary dynamics governed by local interactions \cite{Nahum:2017aa,khemani_operator_2018,von_keyserlingk_operator_2018,nahum_operator_2018,chan_solution_2018,rakovszky_sub-ballistic_2019,garratt_many-body_2020}. Such circuits also form the basis of Google's Sycamore processor \cite{Arute:2019aa}.
The use of minimal models for unitary dynamics is motivated by the fact that analytically tractable models of many-body quantum dynamics remain scarce. While a great deal of understanding has been reached through the study of integrable models, these are non-generic by definition \cite{jstatmech_outofequilibrium_2016,calabrese_introduction_2016,vanicat_integrable_2018,friedman_spectral_2019}. 
Although unitary circuit dynamics exhibit many of the features expected of generic many-body dynamics and present a natural realization of a periodically-driven (Floquet) system \cite{bukov_universal_2015,goldman_periodically_2014}, exact results generally require the presence of randomness in the circuit.

Recently, \emph{dual-unitary circuits} were identified as a class of unitary circuits for which the dynamics of correlations remains tractable, circumventing the need for integrability or randomness \cite{bertini_exact_2019,gopalakrishnan_unitary_2019}. These gates are characterized by the property that the resulting circuit evolution is unitary in both time and space. As a result, correlations vanish everywhere except at the edge of the causal light cone \cite{bertini_exact_2019}, where they can be calculated analytically at all time scales \cite{claeys_maximum_2020}. At long times the resulting correlations can remain constant, oscillating, or decaying, ranging from maximally chaotic to non-ergodic to non-interacting. This makes these models particularly attractive for the study of thermalization: after sufficiently long times, it is expected that all local correlations in a many-body system can be described by a reduced density matrix depending only on the conservation laws present in the system \cite{srednicki_chaos_1994,rigol_thermalization_2008,d2016quantum}.

The study of these models started with the realization that the kicked Ising model (KIM) supported an exact calculation of the spectral form factor and entanglement spectrum at particular values of the coupling constants \cite{bertini_exact_2018,bertini_entanglement_2019}. Ref.~\cite{gopalakrishnan_unitary_2019} subsequently recast the KIM as a unitary circuit and identified dual-unitarity as the underlying reason for the degenerate entanglement spectrum and maximal entanglement growth. Later works have studied more general dynamics, identifying matrix product state initial conditions preserving the solubility of the dynamics \cite{piroli_exact_2019}, correlations of general local operators \cite{gutkin_local_2020}, calculated out-of-time-order correlators \cite{claeys_maximum_2020}, later revisited in the general context of scrambling in random unitary circuits \cite{bertini_scrambling_2020}. Remarkably, the dynamics generated by perturbed dual-unitary gates can still be efficiently described through a path-integral formalism \cite{kos_correlations_2020}.

Almost all such calculations only depend on the dual-unitarity of the underlying gates, such that these models are solvable for any dimension of the local Hilbert space. Despite this salient feature, systematic realizations of dual-unitary circuits remain relatively restricted. Excluding the non-interacting example of a swap-gate, analytical parametrizations of dual-unitary gates are restricted to a $2$-dimensional local Hilbert space (qubits) \cite{bertini_exact_2019} and to kicked models built on complex Hadamard matrices for larger Hilbert spaces \cite{gutkin_local_2020}. Numerically, an iterative protocol has been proposed to generate circuits that are arbitrarily close to dual-unitarity \cite{rather_creating_2019}, but this does not allow analytic predictions or targeting gates with a desired level of ergodicity.

In this work, we present an analytic parametrization of dual-unitary gates for arbitrary local Hilbert space dimension $q$, returning classes of systems for which the dynamics of observables remain analytically tractable. The level of ergodicity of these circuits is classified through the eigenvalues of quantum channels determining the light-cone dynamics, and we show how to systematically realize ensembles of circuits with any desired level of ergodicity and mixing. The steady-state values of the correlation functions are shown to be set by either infinite-temperature Gibbs states or generalized Gibbs ensembles (GGEs) in ergodic and non-ergodic systems respectively. In all examples, the total number of free variables scales as $q^2$, and we show how this encompasses and extends previous parametrizations of dual-unitary gates. Additionally, we illustrate how an ergodicity-inducing perturbation on top of a non-ergodic unitary gate can be introduced without destroying dual-unitarity, leading to a class of solvable models illustrating the appearance of prethermalization to a GGE before eventual thermalization to a featureless infinite-temperature state. 

A Python implementation of all presented calculations is available online \footnote{\href{https://github.com/PieterWClaeys/DualUnitaryCircuits}{https://github.com/PieterWClaeys/DualUnitaryCircuits}}.

\emph{Dual-unitary gates.} We will consider systems where the time evolution is governed by a unitary circuit consisting of two-site operators, where each gate $U$ and its hermitian conjugate can be graphically represented as
\begin{align}\label{eq:def_U_Udag}
U_{ab,cd} = \vcenter{\hbox{\includegraphics[width=0.15\linewidth]{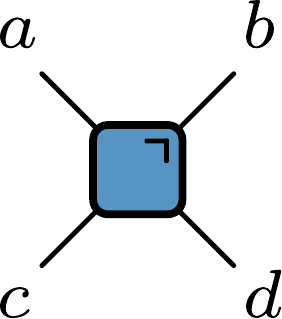}}},\qquad U^{\dagger}_{ab,cd} =  \vcenter{\hbox{\includegraphics[width=0.15\linewidth]{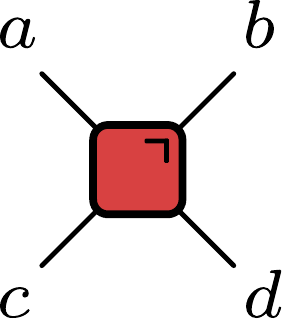}}}. 
\end{align}
In this graphical notation each leg carries a local $q$-dimensional Hilbert space, and the indices of legs connecting two operators are implicitly summed over (see e.g. Ref.~\cite{orus_practical_2014}). Unitarity is graphically represented as
\begin{align}
U U^{\dagger} = U^{\dagger} U = \mathbbm{1} \, \Rightarrow \,\vcenter{\hbox{\includegraphics[width=0.4\linewidth]{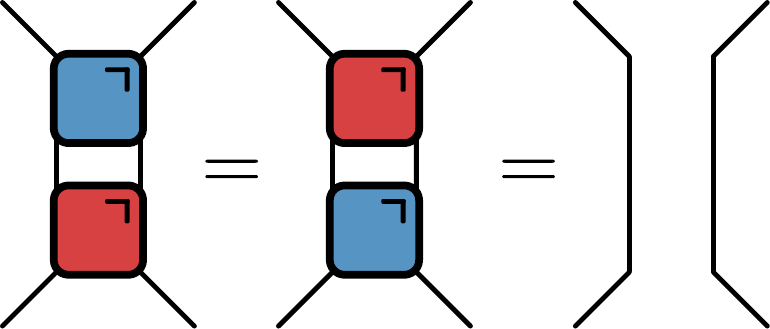}}} \, .
\end{align}
The dual of $U$ is defined through $\tilde{U}_{ab,cd} = U_{db,ca}$, and dual-unitarity is defined as the additional unitarity of $\tilde{U}$ \cite{bertini_exact_2018,gopalakrishnan_unitary_2019},
\begin{align}
\tilde{U} \tilde{U}^{\dagger} = \tilde{U}^{\dagger} \tilde{U} = \mathbbm{1} \,  \Rightarrow \, \vcenter{\hbox{\includegraphics[width=0.51\linewidth]{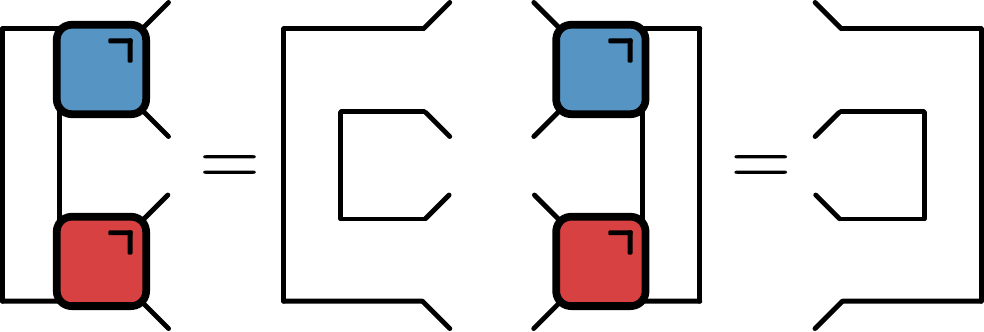}}} \, .
\end{align}
The full evolution $\mathcal{U}(t)$ at time $t$ consists of the $t$-times repeated application of staggered two-site gates
\begin{align}
\mathcal{U}(t) = \vcenter{\hbox{\includegraphics[width=0.8\linewidth]{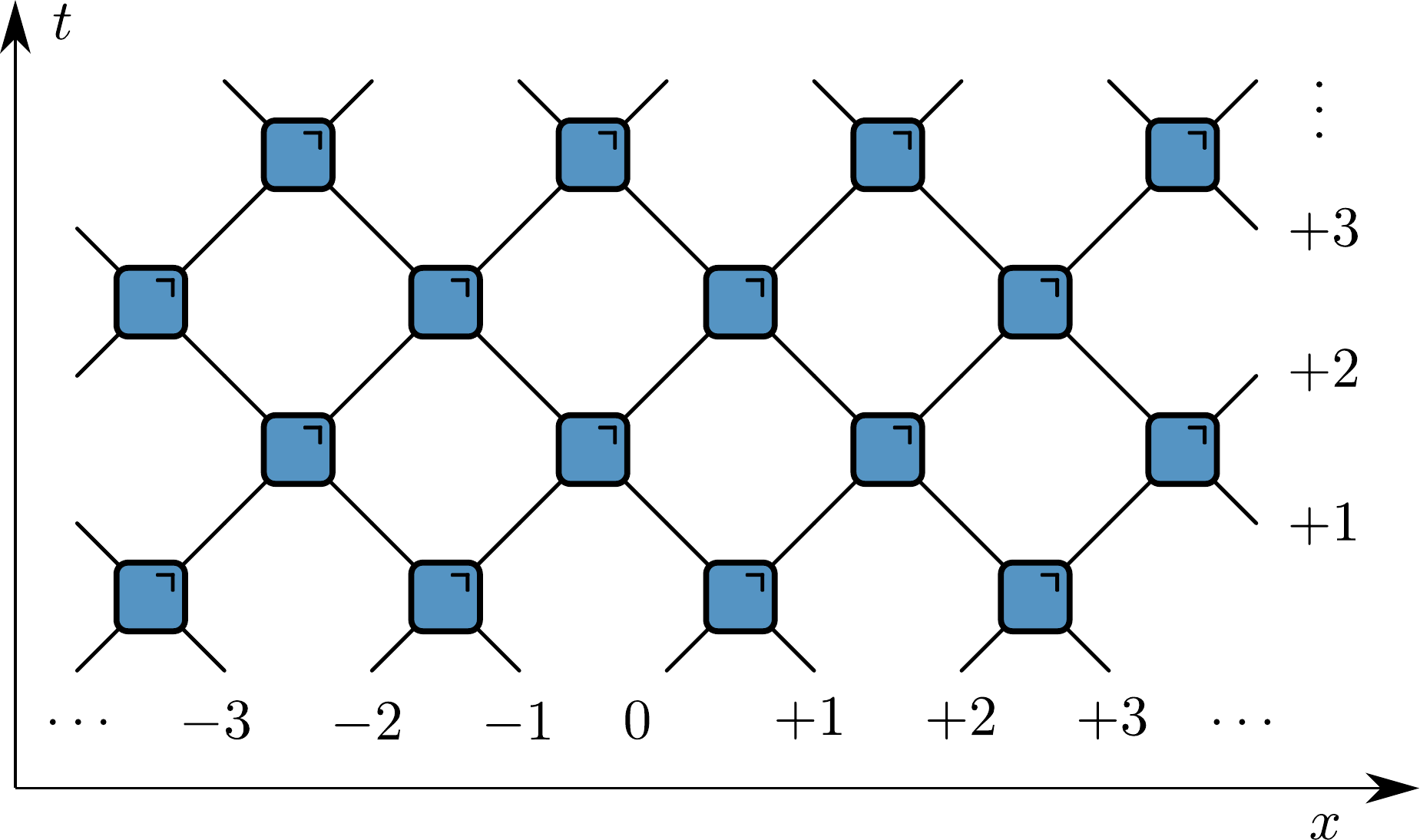}}} \label{eq:def_U_tot}\, . \nonumber
\end{align}
Given an infinite lattice with at each site a local Hilbert space $\mathbbm{C}^q$, we consider correlation functions of the form
\begin{equation}\label{eq:def_corr}
c_{\rho\sigma}(x,t) = \tr\left[ \mathcal{U}^{\dagger}(t)\rho (0) \mathcal{U}(t) \sigma(x) \right]  / \tr(\mathbbm{1}),
\end{equation}
where $\rho, \sigma \in \mathbbm{C}^{q\times q}$ are operators acting on a local $q$-dimensional Hilbert space, and $\rho(x), \sigma(x)$ act as $\rho,\sigma$ on site $x$ and as the identity everywhere else. We will take $\tr(\rho) = 1$ to make the connection with (reduced) density matrices, although this is not a necessary assumption and these can be seen as infinite-temperature correlation functions. Dual-unitarity can be used to show that all correlation functions factorize as $c_{\rho\sigma}(x,t) = \tr(\rho)\tr(\sigma)/q$, except on the edges of the light cone $x=\pm t$ \cite{bertini_exact_2018}. These nontrivial correlation functions can be evaluated as
\begin{align}\label{eq:CorrChannels}
c_{\rho\sigma}(\pm t,t) &= \tr \left[\mathcal{M}_{\pm}^t(\rho)\sigma\right],
\end{align}
where $\mathcal{M}_{\pm} \in \mathbb{C}^{q^2 \times q^2}$ are linear maps defined as
\begin{align}\label{def:channels}
&\mathcal{M}_{+}(\rho) = \tr_{1}\left[U^{\dagger}(\rho \otimes \mathbbm{1})U\right]/q, \\
&\mathcal{M}_{-}(\rho) = \tr_{2}\left[U^{\dagger}(\mathbbm{1} \otimes \rho )U\right]/q.
\end{align}
These are completely positive and trace-preserving maps, acting as a quantum channel. From the unitarity it follows that $\mathcal{M}_{\pm}(\mathbbm{1}) = \mathbbm{1}$, such that these channels are unital. Graphically, we can represent
\begin{align}\label{eq:defM}
\left(\mathcal{M}_{+}\right)_{ab,cd} = \vcenter{\hbox{\includegraphics[width=.2\linewidth]{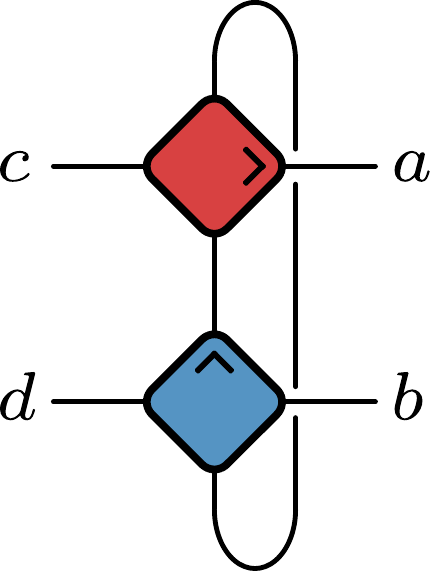}}}, \left(\mathcal{M}_{-}\right)_{ab,cd} = \vcenter{\hbox{\includegraphics[width=.2\linewidth]{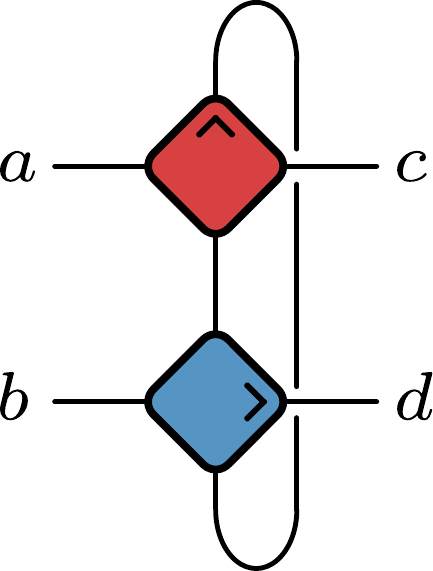}}},
\end{align}
with matrix elements such that $\mathcal{M}(\rho)_{ab} = \sum_{cd} \mathcal{M}_{ab,cd}\rho_{cd}$. Note that light-cone correlation functions can always be calculated in this way, irrespective of dual-unitarity, even in the thermodynamic limit of infinite system size \cite{claeys_maximum_2020}. 

As argued in Ref.~\cite{bertini_exact_2019}, the long-time behaviour of all nontrivial correlations and hence the level of ergodicity is fully determined by the number of eigenvalues $\lambda_{ab}$ of $\mathcal{M}_{\pm}$ with unit modulus, $|\lambda_{ab}| = 1$, with the corresponding eigenoperators acting as non-decaying modes, where we exclude the trivial eigenvalue $1$ corresponding to the identity operator.
At long times, ergodic behaviour is evidenced by the convergence of correlations to their thermal value $\lim_{t\to \infty}c_{\rho\sigma}(\pm t,t) = \tr(\sigma)/q$, consistent with thermalization to an infinite-temperature Gibbs state $\rho_{\textrm{Gibbs}} = \mathbbm{1} / q$ such that $\lim_{t\to \infty}c_{\rho\sigma}(\pm t,t) = \tr(\rho_{\textrm{Gibbs}}\sigma), \forall \sigma$.

The unitary gates are generally not parity-invariant, leading to `chiral' behaviour where $\mathcal{M}_{\pm}$ can have different numbers of nontrivial eigenvalues and corresponding non-decaying modes. In the following, we will only consider the behaviour along $x=t$, governed by the $q^2$ eigenvalues of $\mathcal{M}_+$, but this can be immediately extended to $x=-t$.

\begin{enumerate}
\item \emph{Non-interacting}: All $q^2$ eigenvalues equal 1, and dynamical correlations remain constant.
\item \emph{Non-ergodic}: More than one but less than $q^2$ eigenvalues are equal to 1, dynamical correlations decay to a non-thermal constant.
\item \emph{Ergodic and non-mixing}: All nontrivial eigenvalues are different from 1, but there exists at least one eigenvalue with unit modulus. All correlations oscillate around a time-averaged value corresponding to the thermal value.
\item \emph{Ergodic and mixing}: All nontrivial eigenvalues lie within the unit disc and all dynamical correlations decay to their thermal value.
\end{enumerate}

\newpage

\emph{Parametrization.} We propose a parametrization of dual-unitary gates $U \in \mathbbm{C}^{q^2 \times q^2}$ as
\begin{align}
U =(u_{+}\otimes u_-)V[J](v_- \otimes v_+) = \, \vcenter{\hbox{\includegraphics[width=0.25\linewidth]{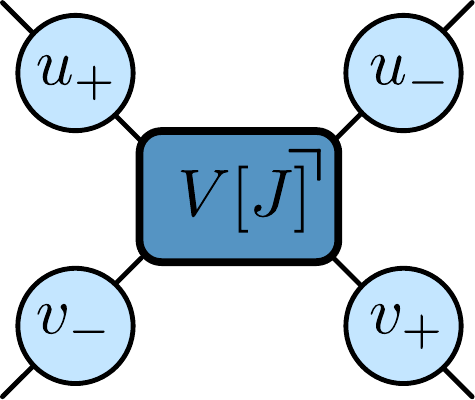}}} \label{eq:param_U}\, ,
\end{align}
with arbitrary one-site unitary gates $u_{\pm}, v_{\pm} \in SU(q)$, and where all entanglement is generated by the two-site unitary $V[J]$ defined as
\begin{equation}\label{eq:VJ}
V[J]_{ab,cd} = \delta_{ad}\delta_{bc}e^{iJ_{ab}},
\end{equation}
with the phases set by an arbitrary real matrix $J \in \mathbbm{R}^{q \times q}$. Both the unitarity and dual-unitarity of $V[J]$ can be readily verified, and the additional one-site unitaries leave both properties intact. We do not expect this parametrization to be exhaustive -- the construction from Ref.~\cite{rather_creating_2019} gives rise to numerically dual-unitary gates that cannot be recast as Eq.~\eqref{eq:param_U}.

Focusing on $x=t$, we write $\mathcal{M}[U] =\mathcal{M}_{+}$ and $u,v = u_{+}, v_{+}$, but calculations for $x=-t$ are completely analogous. Plugging the parametrization \eqref{eq:param_U} in the definition of the quantum channel \eqref{eq:defM}, we find
\begin{align}\label{eq:param_M}
\mathcal{M}[U] &=  (v^{\dagger} \otimes v^T)\mathcal{M}[V](u^{\dagger} \otimes u^T).
\end{align}
The dependence on $u_{-}$ and $v_{-}$ drops out and $\mathcal{M}(J) \equiv \mathcal{M}[V]$ can be evaluated as
\begin{align}
\mathcal{M}(J)_{ab,cd} &= \frac{1}{q}\sum_{e,f=1}^q V[J]^*_{cf,ea} V[J]_{df,eb}  = \sigma_{ab} \delta_{ac}\delta_{bd}.
\end{align}
The channel $\mathcal{M}(J)$ is diagonal, with diagonal elements
\begin{equation}\label{eq:singvalues}
\sigma_{ab} = \frac{1}{q}\sum_{f=1}^q e^{-i(J_{af}-J_{bf})}.
\end{equation}
Written in this way, Eq.~\eqref{eq:param_M} corresponds exactly to a singular value decomposition of $\mathcal{M}[U]$. Graphically, this can be represented as
\begin{equation}\label{eq:expandM}
\mathcal{M} =\vcenter{\hbox{\includegraphics[width=0.30\linewidth]{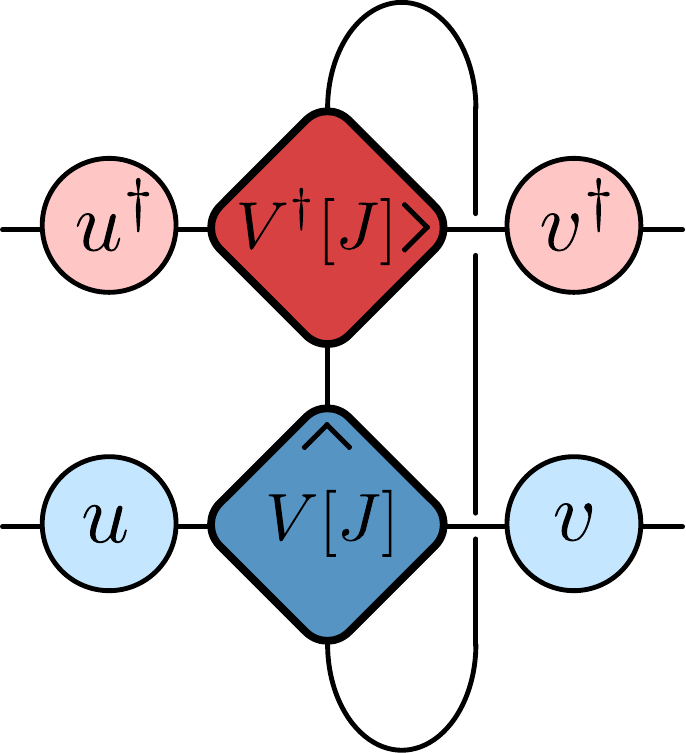}}} = \sum_{ab} \sigma_{ab} \vcenter{\hbox{\includegraphics[width=0.26\linewidth]{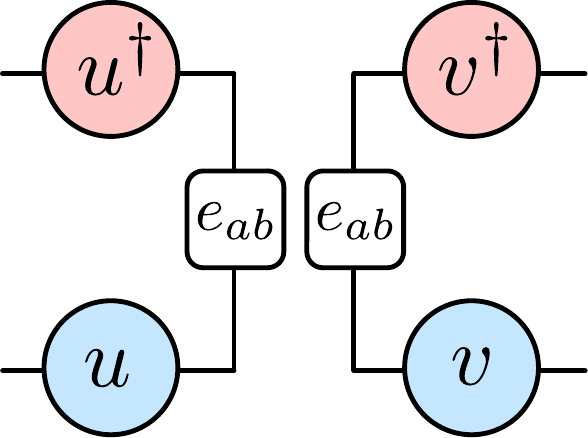}}}.
\end{equation}
The singular values $|\sigma_{ab}| \leq 1$ are the absolute values of the diagonal elements of $\mathcal{M}(J)$, and the unitary transformations are fixed by the one-site unitaries. Furthermore, the left and right eigenoperators of $\mathcal{M}(J)$ are given by the basis operators $e_{ab} \in \mathbbm{R}^{q\times q}$ (defined as the operators with a single non-zero matrix element $ab$) with corresponding eigenvalue $\sigma_{ab}$. 
There are $q$ guaranteed eigenvalues equal to 1, corresponding to the diagonal matrices $e_{aa}, a=1\dots q$, and the remaining $q(q-1)$ eigenvalues arise in complex conjugate pairs, $\sigma_{ab} = \sigma_{ba}^*$. $J$ completely determines the singular values of $U$, such that all quantum channels are guaranteed to have at least $q$ singular values equal to one \footnote{Interestingly, for $q=2$ the demand that the quantum channel has $q=2$ singular values equal to one is equivalent to the demand that the underlying gate is dual-unitary.}. In the following, we will show how the parametrization \eqref{eq:param_U} can be tuned to give rise to classes of dual-unitary models with any given level of ergodicity. The main idea is that $V[J]$ gives rise to a diagonal quantum channel in which we can tune the eigenvalues through $J$, after which the one-site unitaries can be chosen to leave a subset of these eigenvalues invariant.

\emph{Ergodic and mixing gates.} Choosing $J \in \mathbbm{R}^{q \times q}$, $u,v \in SU(q)$ arbitrary, the unitaries do not exhibit any additional structure and the resulting channel will generally only have the trivial eigenvalue associated with the identity. Since $\mathcal{M}[U]$ will generally not be Hermitian, its singular values are unrelated to its eigenvalues, and the left and right eigenoperators will differ. All nontrivial eigenvalues have a modulus smaller than one and $\lim_{t\to \infty} \mathcal{M}^t(\rho) = \tr(\rho) \mathbbm{1}/q = \mathbbm{1}/q$, consistent with thermalization to the infinite-temperature state and thermal correlations $\lim_{t\to \infty}c_{\rho\sigma}(t,t) = \tr(\sigma)/q$. 

\emph{Non-ergodic gates.} Non-ergodic gates have $n$ nontrivial unit eigenvalues with $1\leq n\leq q^2-1$. Such non-ergodic models can be realized in different ways. First, we can consider non-ergodic models where $n \leq q-1$ and the conserved operators are mutually commuting and hence simultaneously diagonalizable. This can be done by imposing a block-diagonal structure on the unitaries $u$ and $v$, effectively turning $n$ singular values into eigenvalues. Taking
\begin{equation}\label{eq:uv_nonerg}
u = w \left[ 
\begin{array}{c|c} 
  \mathbbm{1}_n & 0 \\ 
  \hline 
  0 & u_{q-n} 
\end{array} 
\right] , \qquad v =\left[ 
\begin{array}{c|c} 
  \mathbbm{1}_n & 0 \\ 
  \hline 
  0 & v_{q-n} 
\end{array} 
\right] w^{\dagger},
\end{equation}
in which $u_{q-n},v_{q-n} \in SU(q-n)$ and $w \in SU(q)$, the resulting quantum channel has $n$ additional unit eigenvalues and $n$ mutually commuting eigenoperators given by $c_{a} = w e_{aa} w^{\dagger}, a =1 \dots n$. The block-diagonal matrices preserve the diagonal structure for the eigenvalues $\sigma_{ab}$ of $\mathcal{M}(J)$ with $a,b \leq n$, whereas $w$ leads to a unitary transformation of the quantum channel and its eigenoperators, leaving the eigenvalues invariant \footnote{In this construction $n=q$ is equivalent to $n=q-1$. A $q\times q$ block matrix where one block is the $(q-1) \times (q-1)$ identity matrix is diagonal by construction, and the remaining diagonal element can be set to $1$ by absorbing the phase in $w$. This apparently lost degree of freedom can be associated with the trivial unit eigenvalue by noting that the identity matrix is guaranteed to be conserved and it is impossible to have $q$ diagonal matrices linearly independent from the identity.}.

It follows that $Q_a = \sum_{x \in 2 \mathbbm{N}} c_a(x)$ are conserved quantities for the unitary evolution, satisfying $[Q_a,\mathcal{U}(t=2)] = 0$. The steady-state value of the correlations is determined by the overlap of $\rho$ with the conserved charges. As shown in Appendix, the steady-state values of the correlations are exactly described by a GGE if the initial operator represents a density matrix. In this case, we have
\begin{align}\label{eq:rhoGGE}
\lim_{t \to \infty} \mathcal{M}^t(\rho) = \exp\left[\sum_{a=1}^n (\mu_a-\mu) c_a + \mu \mathbbm{1}\right] = \rho_{\textrm{GGE}},
\end{align}
where the $\mu_a$ and $\mu$ follow from $\rho$ as
\begin{align}
\mu_a = \ln \left(\tr(\rho c_a )\right),\quad \mu =  \ln\left( \frac{1-\sum_{b=1}^n \tr(\rho c_b )}{q-n}\right).
\end{align}
The GGE state necessarily reproduces the correct initial values of all conserved operators, $\tr(\rho c_a) = \tr(\rho_{\textrm{GGE}} c_a), a=1\dots n$, and $\tr(\rho_{GGE}) = 1$, and all correlations decay to the GGE value $\lim_{t \to \infty} c_{\rho\sigma}(t,t) = \tr(\rho_{\textrm{GGE}}\sigma), \forall \sigma$. 
Since we focus on correlations of one-site operators, $\rho_{GGE}$ is a single-site operator corresponding to the reduced density matrix for a single site. However, this does not guarantee that the reduced density matrix for larger subsystems also corresponds to a GGE.

Additional eigenvalues equal to one can be introduced by first introducing additional unit singular values, imposing structure on $J$. However, the additional conserved charges will no longer commute mutually and the steady state can no longer be recast as a GGE. From Eq.~\eqref{eq:singvalues}, a necessary condition for additional unit singular values is for multiple rows of $J$ to be equal. Taking the first $m<n$ rows of $J$ to be equal leads to $m(m-1)$ additional singular values $\lambda_{ab} = 1, a, b \leq m$. For $m \leq n$ the block-diagonal structure of of Eq.~\eqref{eq:uv_nonerg} again leaves these singular values invariant, and the total gate has additional unit-eigenvalue eigenoperators and hence conserved charges $w e_{ab} w^\dagger,  a,b \leq m$. While the final value is no longer described by a GGE, it will still converge to a non-thermal value set by the overlap of $\rho$ with the (properly orthonormalized) conserved charges.
This can be seen as the limit of non-mixing behaviour: taking rows of $J$ to be equal up to a non-zero constant, e.g. for fixed $a,b \leq n$ setting $J_{af} = J_{bf} + \phi, f = 1\dots q$, leads to a pair of complex conjugate eigenvalues $\lambda_{ab} = (\lambda_{ba})^* = e^{i\phi}$. The resulting correlation functions do not decay but exhibit persistent oscillations $\propto e^{i \phi t}$, averaging out to zero for non-zero $\phi$, such that the time-average value corresponds to the GGE value~\eqref{eq:rhoGGE} in the absence of equal rows.

\emph{Non-interacting models.} Non-interacting models are characterized by all eigenvalues equal to one. This can be done by first setting all rows of $J$ to be equal, such that all singular values are equal to 1. In this case $V[J]$ corresponds to a swap gate and $\mathcal{M}(J) = \mathbbm{1}$, such that $\mathcal{M}[U] = v^{\dagger} u^{\dagger} \otimes vu $. All eigenvalues have modulus one, where all eigenvalue are exactly one if $v = u^{\dagger}$. All dynamical correlations remain constant and $\mathcal{M}[U] = \mathbbm{1}$.

\emph{Ergodic and non-mixing.} As a final example, ergodic but non-mixing gates are characterized by $1 \leq n \leq q^2-1$ nontrivial eigenvalues that are all different from one but with unit modulus. This can be done for generic $J$ by setting $u = w P,\ v = w^{\dagger}$, with $w \in SU(q)$ and in which $P$ is defined as $P_{a,b} = e^{i \theta_a} \delta_{a,a+1}$, identifying $q+1 \equiv 1$, and $\theta_{a}, a =1\dots q$ are arbitrary phases. Considering the subspace of all unit-eigenvalue (diagonal) eigenoperators of $\mathcal{M}[V]$, the effect of $P$ is to set $P^{\dagger}e_{aa} P = e_{a+1,a+1}$, such that $\mathcal{M}(w^{\dagger} e_{aa} w) = w^{\dagger} e_{a+1,a+1} w$. Within this degenerate subspace, $\mathcal{M}$ acts as a shift operator, for which the known eigenvalues are given by $e^{2\pi i f/q}, f=1\dots q$, where the trivial eigenvalue $1$ corresponds to the identity. As such, this leads to $q-1$ nontrivial eigenvalues given by the remaining roots of unity. At sufficiently long times, all correlation functions remain nonzero and oscillate around the zero ergodic value, satisfying $c_{\rho \sigma}(t+q,t+q) = c_{\rho \sigma}(t,t)$. This effectively realizes a discrete time crystal, where the correlations in a periodically-driven system respond with a period that is an integer multiple of the driving period \cite{else_floquet_2016,khemani_phase_2016,yao_discrete_2017}.

\begin{figure}[tb!]
\includegraphics[width=\columnwidth]{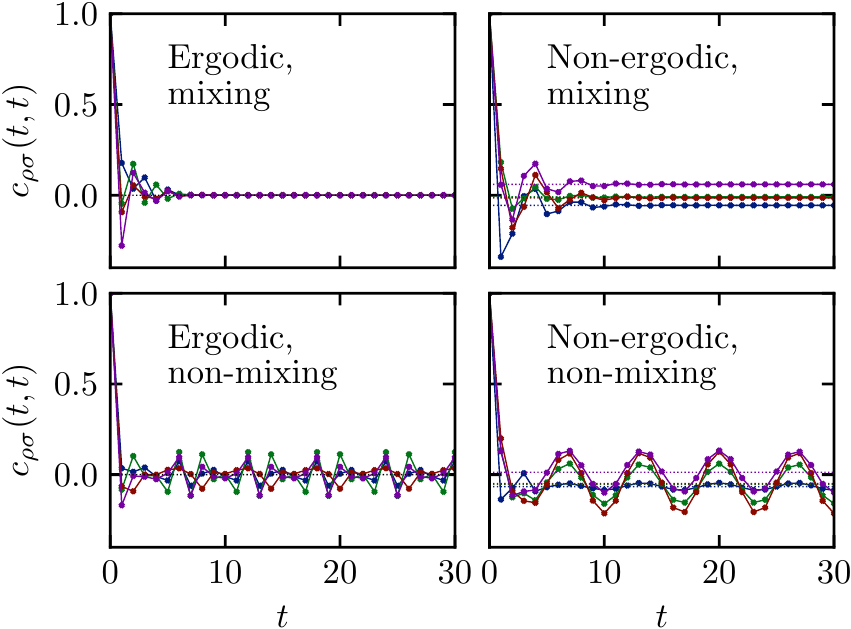}
\caption{Evolution of correlation functions $c_{\rho \sigma}(t,t)$, where $\rho,\sigma \in \mathbbm{C}^{q \times q}$ are randomly generated matrices with $\tr(\sigma)=0$ leading to a thermal value $c_{\rho \sigma}(t,t) \to 0$. Local Hilbert space dimension $q=6$ and $4$ different operators are considered. After an initial transient regime, in the ergodic models the correlations either exponentially decay to zero (mixing) or oscillate around zero (non-mixing) with period $q$, whereas in the non-ergodic models the correlations decay to a non-zero value (mixing) with possible oscillations around these non-zero values with a tunable period (non-mixing).
\label{fig:correlations}}
\end{figure}

\emph{Examples.} -- In Fig.~\ref{fig:correlations}, we present numerical examples for different dynamics. Note that the quantum channel construction does not require all unitaries to be identical \cite{bertini_exact_2018}, such that individual unitary gates can be randomly selected while still keeping the overall level of ergodicity of the full circuit. As shown in Appendix, the level spacing statistics can also be calculated,  a common indicator of chaos and ergodicity, returning the expected GUE statistics for ergodic and mixing gates and Poisson statistics otherwise, consistent with the proposed classification \cite{atas_distribution_2013}. 

As an additional example, we consider a dual-unitary model for prethermalization with an arbitrarily large local Hilbert space. For a non-ergodic model, the system locally thermalizes to a GGE consistent with the conserved charges. Any perturbation generally destroys all nontrivial conservation laws, inducing thermalization to the infinite-temperature state. However, for small perturbations we expect a separation of time scales: the correlations initially prethermalize to the GGE values of the non-ergodic model before an eventual thermalization to the infinite-temperature thermal values \cite{bertini_prethermalization_2015,mori_thermalization_2018}. 

\begin{figure}[tb!]
\includegraphics[width=\columnwidth]{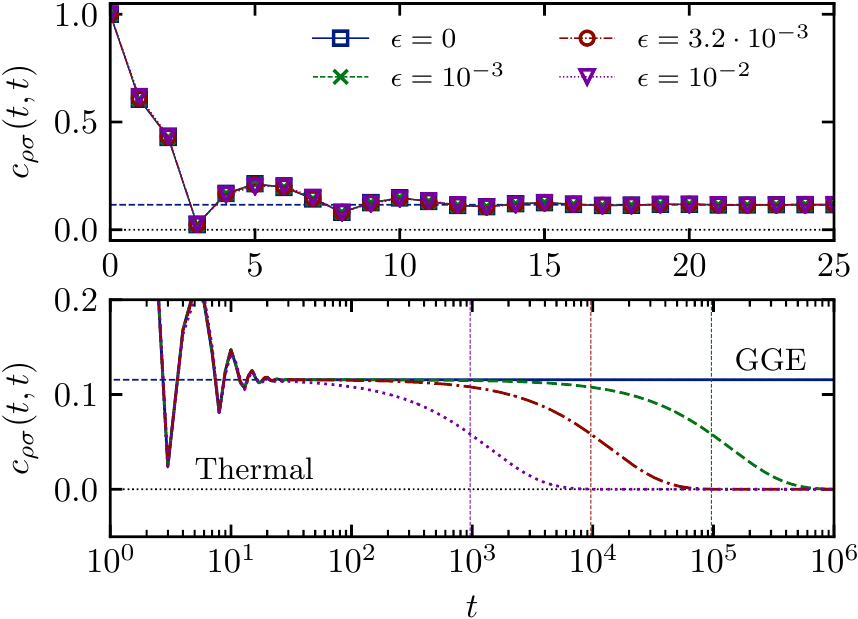}
\caption{Evolution of correlation function $c_{\rho\sigma}(t,t)$ at short and long time scales for a randomly-generated traceless $\sigma$ and a randomly-generated density matrix $\rho$ with $\tr(\rho\sigma) = 1$ for a non-ergodic gate $U$ on top of which an ergodicity-inducing perturbation is added with strength $\epsilon$. Local Hilbert space dimension $q=6$. Vertical dashed lines mark $t = \log(2)/(1-|\lambda|) \propto \epsilon^{-2}$, with $\lambda$ the dominant nontrivial eigenvalue of $\mathcal{M}$, and horizontal lines denote the thermal and GGE values. 
\label{fig:prethermalization}}
\end{figure}

Given general $J$, we can introduce a small ergodicity-inducing perturbation on top of a non-ergodic model starting from Eq.~\eqref{eq:param_M}, setting
\begin{align*}
u =  e^{i \epsilon W_u}  w \left[ 
\begin{array}{c|c} 
  \mathbbm{1}_n & 0 \\ 
  \hline 
  0 & u_{q-n} 
\end{array} 
\right],\,
v = \left[ 
\begin{array}{c|c} 
  \mathbbm{1}_n & 0 \\ 
  \hline 
  0 & v_{q-n} 
\end{array} 
\right]w^{\dagger}  e^{-i \epsilon W_v},
\end{align*}
with again $u_{q-n},v_{q-n} \in SU(q-n)$, $w \in SU(q)$, and where the perturbation is generated by two (non-equal) Hermitian operators $W_{u,v} \in \mathbbm{C}^{q\times q}$ and can be tuned through $\epsilon$. At $\epsilon=0$, this reduces to a non-ergodic model with $n$ conservation laws, whereas any finite $\epsilon$ results in an ergodic model with all nontrivial eigenvalues of the quantum channel having a modulus smaller than one. This is illustrated in Fig.~\ref{fig:prethermalization}, where for small $\epsilon$ the dynamics of the different circuits are indistinguishable, seemingly converging to the non-thermal steady-state value of the non-ergodic model. However, the effect of the perturbation becomes apparent at longer times, where the models with nonzero $\epsilon$ eventually thermalize to the infinite-temperature state indicated by vanishing correlations. The time scale needed to reach the eventual thermal state is fully determined by the subleading eigenvalue of $\mathcal{M}$ and scales as $\epsilon^{-2}$, as it can be verified from degenerate perturbation theory that the first-order correction on the unit eigenvalues vanishes. 

\emph{Conclusion.} -- It was shown how to generate classes of dual-unitary or maximally-entangled operators with arbitrary local Hilbert space dimension and any desired level of ergodicity. Evolving a local operator under a circuit composed of dual-unitary gates, the dynamics of local correlations remains analytically tractable for any local Hilbert space dimension and without the need for randomness, such that these models can be used to study both chaotic and non-ergodic dynamics in systems with an arbitrarily large Hilbert space. Focusing on one-site operators, the steady-state correlations were analytically shown to be given by the infinite-temperature Gibbs state (ergodic) or a generalized Gibbs ensemble (non-ergodic), where we also illustrated prethermalization to the latter before eventual thermalization to the former in a non-ergodic model with added ergodicity-inducing perturbation.

\emph{Acknowledgements.} -- We gratefully acknowledge support from EPSRC Grant No. EP/P034616/1. 

\appendix

\section*{Appendix}

\emph{Connection with known parametrization.} Currently, two classes of dual-unitary models are known. For arbitrary local Hilbert space dimension, a family of kicked models was recently introduced building on complex Hadamard matrices \cite{gutkin_local_2020}. An explicit parametrization of the gates and their eigenvalues and eigenvectors was given for so-called Discrete Fourier Transform chains. Following the results of Ref.~\cite{gopalakrishnan_unitary_2019} for the KIM, the underlying gates can be rewritten as (up to diagonal one-site unitaries on the outer legs, which do not influence dual-unitarity),
\begin{align}
U = \mathcal{I} (\mathcal{K} \otimes \mathcal{K}) \mathcal{I}  = \vcenter{\hbox{\includegraphics[width=0.16\linewidth]{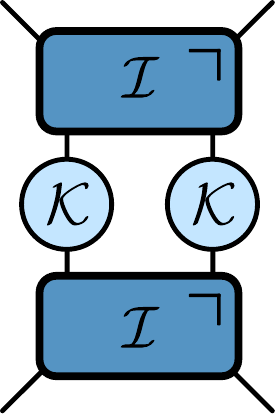}}}\, .
\end{align}
The one-site and two-site unitaries $\mathcal{K}$ and $\mathcal{I}$ are defined as $\mathcal{K}_{ab} =  e^{2\pi i ab/q} / \sqrt{q}$ and $\mathcal{I}_{ab,cd} = \delta_{ac}\delta_{bd} e^{2\pi i ab/q}$ respectively. The matrix elements of $U$ can be evaluated as
\begin{equation}
U_{ab,cd} = \frac{1}{q}\exp\left[\frac{2 \pi i}{q} (a+d)(b+c)\right],
\end{equation}
which can be recast as Eq.~\eqref{eq:param_U} by writing $U = (\mathcal{K}^{\dagger} \otimes \mathcal{K}) V[J] (\mathcal{K} \otimes \mathcal{K}^{\dagger})$, with $V[J]$ defined as in Eq.~\eqref{eq:VJ} with a factorizable $J_{ab} = 2 \pi a b /q$. We also note that Eq.~\eqref{eq:singvalues} implies that the singular values of the corresponding quantum channel immediately follow as
\begin{equation}
|\sigma_{ab}| = \frac{1}{q}\left|\sum_{f = 1}^q e^{-2 \pi i  (a-b)f /q}\right| = \delta_{ab}.
\end{equation}
Such models are characterized by quantum channels with $q$ unit singular values and $q(q-1)$ zero singular values. Since the possible singular values of these quantum channels are much more restricted than those of Eq.~\eqref{eq:param_M}, these unitaries are necessarily a subclass of the proposed parametrization.

The second class consists of dual-unitary gates for $q=2$ \cite{bertini_exact_2019}. Up to a global phase, Eq.~\eqref{eq:param_U} provides the most general parametrization of dual-unitary gates with 
\begin{align}
V[J] = &\exp\left[-i\frac{\pi}{4}\left(\sigma^x \otimes \sigma^x + \sigma^y \otimes \sigma^y\right)\right] \nonumber\\
&  \qquad \times  \exp\left[-iJ \sigma^z \otimes \sigma^z\right],
\end{align}
in which $J \in \mathbbm{R}$ is a continuous variable. Its matrix elements can be straightforwardly evaluated as
\begin{equation}\label{eq:Vqubit}
V[J]_{ab,cd} = \delta_{ac}\delta_{bd} \exp\left[i \left((2 J  - \pi/2) (a-b)^2-J\right)\right], \nonumber
\end{equation}
corresponding exactly to the proposed parametrization \eqref{eq:param_U}. The singular values follow from Eq.~\eqref{eq:singvalues} as $\sigma_{11}=\sigma_{22} = 1$ and $\sigma_{12} = \sigma_{21} = \sin(2J)$. Choosing the rows to be equal requires $J = \pi/4$, and $V[J = \pi/4]$ is a swap gate up to a constant phase $e^{-i\pi/4}$. Choosing $J=0$ sets all nontrivial singular values equal to zero and the phase can be rewritten as $e^{-i \pi a^2/2} e^{-i \pi b^2/2} e^{-i \pi ab}$. The first two phases can be absorbed in diagonal one-site unitaries, whereas the final phase corresponds to the Discrete Fourier Transform parametrization, consistent with known results for the KIM \cite{gutkin_local_2020}. While there are four possible phases $J_{ab}$ in Eq.~\eqref{eq:param_U} for $q=2$, one can be eliminated through the global phase, and the other two can be eliminated by one-site unitaries on the legs, such that Eqs.~\eqref{eq:Vqubit} and \eqref{eq:param_U} correspond.  

\emph{Calculation of steady-state values}. -- Taking $\rho$ to be a generic initial operator and $U$ a non-ergodic dual-unitary circuit with $n < q$ conservation laws parametrized as in Eq.~\eqref{eq:uv_nonerg}, the steady-state values can be explicitly calculated, and furthermore expressed as a Generalized Gibbs Ensemble in the case where $\rho$ is a density matrix. The eigenoperators of $\mathcal{M}$ are given by $\{c_{a} = w e_{aa} w^{\dagger}, a=1\dots n\}$, which can be seen as projection operators on $w\ket{a}$ with $\ket{a}_b = \delta_{ab}$, and the identity matrix $\mathbbm{1}$. The former are trace-orthonormal amongst themselves, but have a non-zero overlap with the identity matrix. A trace-orthonormal set of hermitian operators can be found setting $c_{n+1} = (\mathbbm{1} - \sum_{a=1}^n c_{a})/\sqrt{q-n}$, satisfying $\tr(c^{\dagger}_a c_b) = \tr(c_a c_b) = \delta_{ab}$. 

For any initial operator $\sigma$, its non-decaying part under the action of $\mathcal{M}$ follows as
\begin{align}
\lim_{t\to\infty}\mathcal{M}^t(\sigma) = \sum_{a=1}^{n+1} \tr(c_a \sigma) c_a.
\end{align}
Considering the special case where $\sigma=\rho$ is a one-site density matrix, $\tr(\rho)=1$, and positive semi-definite such that $0 \leq \tr(c_a \rho)  = \langle a|w^{\dagger} \rho w|a\rangle \leq 1, a=1\dots n$, this steady-state can be expressed as a GGE. Expanding $c_{n+1}$ and making use of $\tr(\rho)=1$, we find
\begin{align}
\lim_{t\to\infty}\mathcal{M}^t(\rho) = &\sum_{a=1}^n \tr(c_a \rho) c_a \nonumber\\
& + \frac{1 - \sum_{b=1}^n \tr(c_b \rho)}{q-n}\left(\mathbbm{1}-\sum_{a=1}^n c_a\right).
\end{align}
This can be recast as a GGE
\begin{align}
\rho_{GGE} = \exp\left[\sum_{a=1}^n (\mu_a - \mu ) c_a + \mu \, \mathbbm{1}\right],
\end{align}
since all involved operators commute and are diagonalized by $w$. In this diagonal basis we can make use of
\begin{align}
&\exp\left[\sum_{a=1}^q \lambda_a e_{aa}\right] =\sum_{a=1}^q \exp[{\lambda_a}] e_{aa},
\end{align}
to write
\begin{align}
\rho_{GGE} = \sum_{a=1}^n e^{\mu_a} c_a + e^{\mu} \left(\mathbbm{1}-\sum_{b=1}^n c_b\right).
\end{align}
Identifying both expressions returns $\mu_a = \ln \left(\tr(c_a \rho)\right), a = 1 \dots n, $ and 
\begin{align}
\mu = \ln\left( \frac{1-\sum_{b=1}^n \tr(c_b \rho)}{q-n}\right).
\end{align}
Note that this mapping to real parameters $\mu_a$ depends on $\rho$ being a density matrix, since otherwise the arguments of the logarithms are allowed to be negative. The final $\rho_{GGE}$ satisfies $\tr(\rho_{GGE}) = 1$ and $\tr(\rho_{GGE}c_a) = \tr(\rho c_a), a=1\dots n$ and captures all nontrivial steady-state correlations since $\lim_{t \to \infty}\tr\left[ \mathcal{M}^t(\rho) \sigma\right] = \tr(\rho_{GGE} \sigma), \forall \sigma \in \mathbbm{C}^{q \times q}$. 

The steady-state value for additional non-commuting conservation laws obtained by setting the first $m$ rows of $J$ to be equal can be found by extending the trace-orthonormal basis with $c_{ab} = w(e_{ab}+e_{ba})w^{\dagger}/\sqrt{2}, b<a \leq m$ and $c_{ab} = w(e_{ab}-e_{ba})w^{\dagger}/\sqrt{2}, a < b \leq m$. This leads to
\begin{align}
\lim_{t\to\infty}\mathcal{M}^t(\rho) &= \sum_{a=1}^n \tr(c_a \rho ) c_a
+\sum_{a=1}^m \sum_{b \neq a}^m \tr(c_{ab}^\dagger \rho)c_{ab} \nonumber\\
& + \frac{1 - \sum_{b=1}^n \tr(c_b \rho)}{q-n} \left(\mathbbm{1}-\sum_{a=1}^n c_a \right) .
\end{align}

\emph{Level statistics.} Given the total unitary evolution operator over a single period, the statistic properties of its eigenvalues are a common indicator of ergodic or chaotic behaviour \cite{atas_distribution_2013,d2016quantum}. Labeling the eigenspectrum of this unitary operator as $e^{i\theta_n}$, $\theta_n < \theta_{n+1}$, the average level spacing ratio is defined as the average value of
\begin{align}
r_n = \frac{\textrm{min}(s_n,s_{n+1})}{\textrm{max}(s_n,s_{n+1})}, \qquad s_n = \theta_{n+1}-\theta_{n}.
\end{align}
For an ergodic circuit we expect $\langle r \rangle = 0.603$, the Gaussian Unitary Ensemble (GUE) value indicating random matrix statistics, whereas a non-ergodic circuit should lead to Poissonian statistics and $\langle r \rangle = 0.386$ \cite{atas_distribution_2013}.

Numerically, we calculated the level spacing ratio for a single realization of the evolution operator and $q=2,3,4$, consisting of $2$ layers of $3$ gates acting on $6$ local sites with periodic boundary conditions. Each individual gate is randomly generated from the appropriate ensemble, breaking translational invariance. For all considered choices of $q$,  the ergodic and mixing gates return $\langle r \rangle$ close to the GUE value, whereas all other classes lead to $\langle r \rangle$ close to the Poisson value.

\begin{figure}[tb!]
\includegraphics[width=\columnwidth]{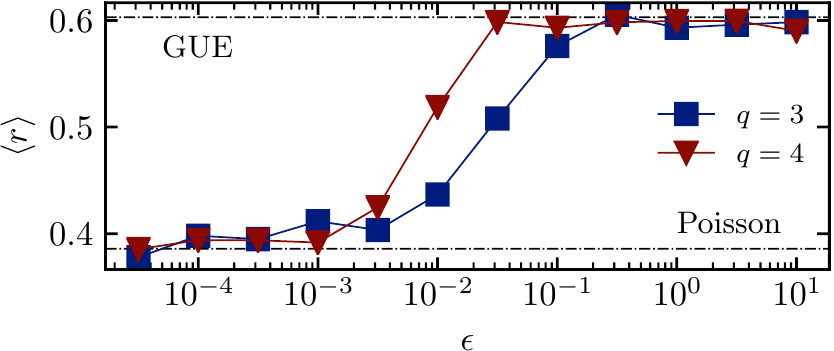}
\caption{Average level spacing ratio $\langle r \rangle$ for increasing perturbation strength $\epsilon$ given an ergodicity-inducing perturbation on top of a non-ergodic model with $q=3,4$ and $n=1$ and a finite system acting on $6$ sites. Horizontal lines denote the Poisson and GUE values.
\label{fig:levelspacingratio}}
\end{figure}

This is also illustrated for the model for prethermalization in Fig.~\ref{fig:levelspacingratio}, again using the same sampling method. For finite systems the level spacing ratio returns the non-ergodic Poisson value at sufficiently small perturbation strength $\epsilon$, whereas GUE statistics arise at large perturbation strength. The crossover occurs at smaller $\epsilon$ for larger $q$.

\bibliography{Library.bib}
\end{document}